\documentstyle {caosp}
\input epsf.sty
\begin{document}
\title{The energy distribution of $\beta$ CrB for the specific 
stellar abundances}
\author{F. Castelli}
\institute {CNR-Gruppo Nazionale Astronomia and Osservatorio Astronomico
di Trieste}
\maketitle
\begin{abstract}
The comparison of the observed and computed energy distributions 
of $\beta$ CrB 
has shown that a model with the specific chemical composition 
of the star can account for the visual enery distribution, while it is
still unable to reproduce ultraviolet observations shortward of 1700 \AA.
Furthermore, the predicted absorption of strong Fe\,{\sc ii} and
Mg\,{\sc ii} UV lines is much larger than the observed one.
\end{abstract}
\section{Introduction}
Some magnetic Ap stars, as  $\beta$ CrB (HD 137909, HR 5747, F0p) and 
33 Lib (HD 137949) show an excess of ultraviolet flux 
shortward of 2000~\AA~ when compared both  with other peculiar or normal stars 
of similar spectral type (i.e. 60 Tau = HD 27628, 78 Tau = HD 28319) and with 
energy distributions computed for solar or solar scaled abundances.

Hack et al. (1997) explained the UV excess of $\beta$ CrB
($T_{\rm eff} = 7950$~K, $\log g = 4.3$ from their paper) as due to 
a $\lambda$ Boo companion having [M/H]=-1.0 and $T_{\rm eff}$ equal to about 
8200~K. However, the hypothesis of a silicon deficiency which increases
the ultraviolet flux was also suggested.

In this paper, we show that models computed with approximate specific 
abundances of $\beta$~CrB explain the observed depressions at 4200~\AA~
and 5200~\AA, but are not able to explain the observed UV excess.

\section{The far UV continous absorptions}

Figure~1 shows that
the energy distribution of stars with $T_{\rm eff} = 8000$~K and $\log g = 4.0$
is dominated shortward of 1700~\AA~  by the discontinuities
of C\,{\sc i} (1240~\AA; 1445~\AA) and of 
Si\,{\sc i} (1520~\AA~; 1677~\AA).  
The red wing of Ly$_{\alpha}$ yields also an important contribution up to 
about 1450~\AA~ in  stars with solar metallicity, but its effect on the
observed flux decreases with increasing metallicity.

Therefore, when the energy distribution is computed, correct values for the
carbon and silicon abundances should be used. 
While the ATLAS9 code (Kurucz, 1993) does not
permit to change the abundances in an arbitrary way, owing to the 
pretabulated ODF's functions, arbitrary abundances can be used
in the ATLAS12 code (Kurucz, 1996), which computes line blanketing by 
means of the opacity sampling method.
\begin{figure}[hbt]
\epsfysize=11cm
\epsfxsize=5.0cm
\hspace{1.0cm}\epsfbox{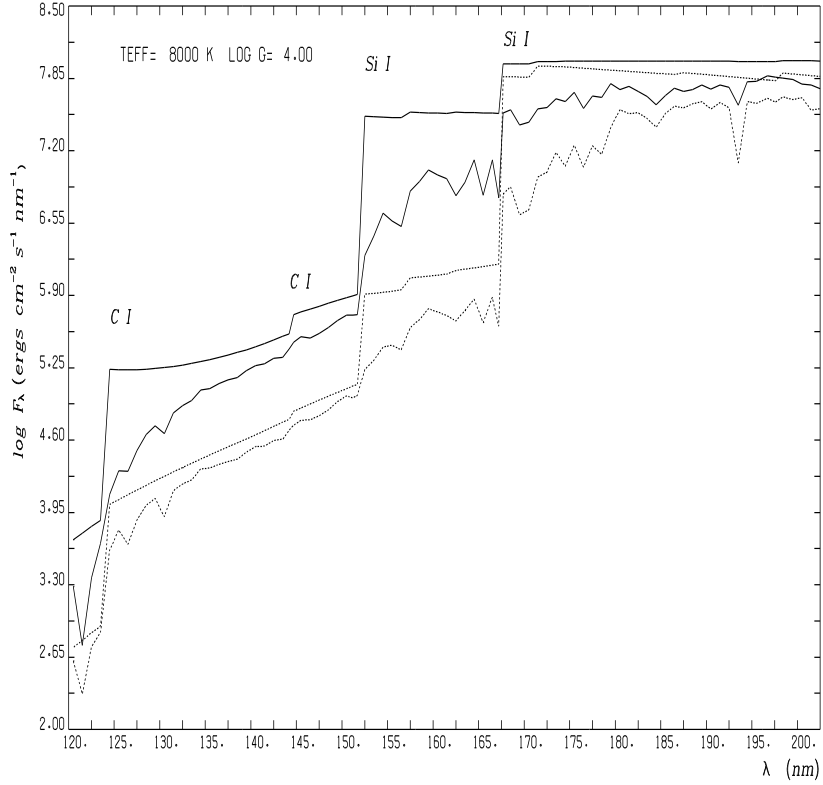}
\caption[h]{Comparison of fluxes from models having the same parameters 
$T_{\rm eff} = 8500$~K and $\log g = 4.0$
and different metallicities [M/H]=0.0 (full line), and [M/H]=+1.0 (dotted
line).
The continous absorption is also plotted for each model} 
\end{figure}

\section{The computed energy distribution for specific abundances}

To derive the energy distribution for the specific abundances of $\beta$ CrB,
we performed the following steps:

1) By comparing the dereddened observed c and (b-y) indices with the
computed ones we derived
the model parameters $T_{\rm eff} = 7950$~K, $\log g = 4.3$, and by comparing 
the observed and computed magnitudes (relative to 5556~\AA) in the Balmer jump 
region we derived [M/H]=-1.0. 
Observed spectrophotometry was taken from Pyper \& Adelman (1985).
  
2) We compared observed and computed high-resolution spectra in the visible
region. The abundances and the microturbulent velocity were modified to
fit the largest number of observed features.
Spectroscopic observations cover the ranges 4130-4544~\AA~ at 20000 resolution,
6076-6163~\AA~ and 6629-6720~\AA~ at 45000 resolution. They were taken at 
Haute-Provence Observatory by R. Faraggiana and at Crimean Astrophysical 
Observatory by N. Polosukhina. The synthetic spectra were computed by using 
the SYNTHE code (Kurucz, 1993) and the atomic line data from Kurucz (1995).
The abundances listed in Table~1 correspond to
a microturbulent velocity $\xi = 0$~km\,s$^{-1}$. While the spectra are better 
fitted by $\xi=2$~km\,s$^{-1}$ for $\lambda > 6000$~\AA~,
the zero value yields very strong computed features in the 4300~\AA~ region.
Because most of the lines are so weak for $\lambda$$>$ 6000~\AA~ that 
the differences between spectra computed with different $\xi$ values are not 
very large,  we preferred to adopt a zero value
for $\xi$. Very probably the inconsistency between the line strengths
at 4300~\AA~ and 6000~\AA~ would be  removed with the
inclusion of the magnetic field effects in the synthetic spectrum.

Table~1 shows that the estimated abundances of $\beta$~CrB are very 
different from solar scaled abundances. In fact, the iron abundance is 
10 times the solar one, but the silicon abundance is solar, 
and the carbon abundance is less than solar. 
Finally, the abundances of the heavy elements are enhanced much more than 10 
times over the solar ones. 

3) By using the abundances listed in Table~1 
we computed an ATLAS12 model and a whole synthetic spectrum at 500000
resolution
for the range 900-8000 ~\AA. Then, we degraded the computed spectrum at
the ODF's resolution, at 3~\AA~ resolution, and at 1~\AA~ resolution, in
order to compare it with the ATLAS9 energy distribution and with IUE
observations.

\begin{table}[t]
\small
\begin{center}
\caption{The abundances $\log (N_{\rm elem}/N_{\rm tot})$ for $\beta$ CrB
for $T_{\rm eff} = 7950$~K, $\log g = 4.3$, [M/H]=-1.0 and
$\xi = 0$~km\,s$^{-1}$}
\label{t1}
\begin{tabular}{lllll|lllll}
\hline\hline
&Elem&~~Sun&$\beta$ CrB&~[M/H]&&Elem&~~Sun&$\beta$ CrB&~[M/H]\\
\hline
~3&Li&-10.88&-8.28&[$+2.60$]&25&Mn&$~-6.65$&-5.65&[+1.00]\\
~6& C&~-3.48 &-3.98&[$-0.50$]&26&Fe&$~-4.53$&-3.27&[+1.26]\\
~7& N&~-3.99 &-3.99&[$+0.00$]&27&Co&$~-7.12$&-6.12&[+1.00]\\
~8& O&~-3.11 &-4.00&[$-0.89$]&28&Ni&$~-5.79$&-5.69&[$+0.00$]\\
11&Na&~-5.71&-5.71&[$+0.00$]&31&Ga&$~-9.16$&-7.66&[+1.50]\\
12&Mg&~-4.46&-4.46&[$+0.00$]&56&Ba&$~-9.91$&-7.91&[+2.00]\\
13&Al&~-5.57&-6.77&[$-1.20$]&57&La&$-10.82$&-8.17&[+2.65]\\
14&Si&~-4.49&-4.49&[$+0.00$]&58&Ce&$-10.49$&-7.09&[+3.40]\\
20&Ca&~-5.68&-5.00&[$+0.68$]&62&Sm&$-11.04$&-9.04&[+2.00]\\
21&Sc&~-8.94&-7.94&[$+1.00$]&63&Eu&$-11.53$&-6.63&[+4.90]\\
22&Ti&~-7.05&-6.05&[$+1.00$]&64&Gd&$-10.92$&-7.02&[+3.90]\\
23& V&~-8.04&-7.04&[$+1.00$]&70&Yb&$-10.96$&-8.96&[+2.00]\\
24&Cr&~-6.37&-4.67&[$+1.70$]\\
\hline\hline
\end{tabular}
\end{center}
\end{table}

\section {Results}
Figure~2 shows that the new computed energy distribution reproduces 
very well the depressions observed at 4200~\AA, and rather well that 
observed at 5200~\AA. Our computations indicate that
these depressions are due to the very large number of lines of heavy elements 
observed in these regions, in particular Ce and Gd. 

\begin{figure}[hbt]
\epsfysize=11cm
\epsfxsize=5.0cm
\hspace{0cm}\epsfbox{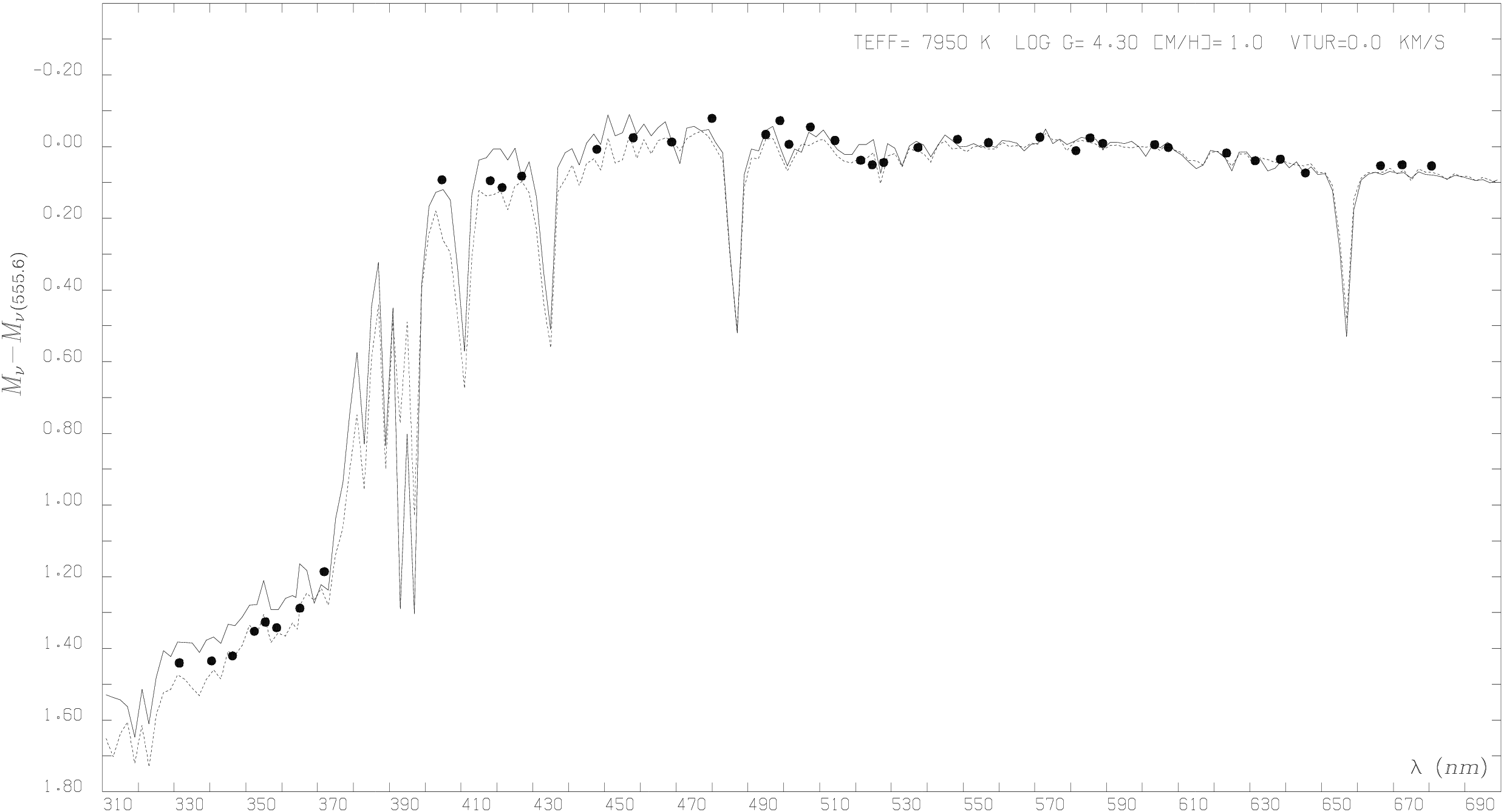}
\vspace{0.5cm}
\caption[h]{The visible energy distributions (in mag, relative to 5556~\AA)
from both ATLAS9 (full line) and ATLAS12 (dotted line) models are 
compared with the observations (dark points).}
\end{figure}

Figure~3 compares the observed and computed energy distributions 
in the ultraviolet region. The data of the IUE image SWP44998 (dashed line) 
well correspond to the observations taken from the TD1 S2/68 Catalogue
(Jamar et al., 1976) (points). 
The excess of observed ultraviolet flux con not be predicted
by a model computed with the specific abundances of the star. Also the
computed strong blanketing at 2300~\AA~ is not observed. The analysis of the
high-resolution IUE image LWR07000 has shown that all the strong Fe\,{\sc ii}
lines of the UV multiplets 2 and 3 are much weaker than the predicted ones.
 The discrepancy is not removed even by assuming solar iron abundance 
for $\beta$ CrB. This behaviour is common to all the strong ultraviolet lines,
including the Mg\,{\sc ii} lines at 2800~\AA.   
Babel (1994) observed a similar disagreement for the Ca\,{\sc ii} resonance
lines and he explained it with abundance stratifications.

\section{Conclusions}

An atmospheric model computed for the specific abundances of $\beta$~CrB
is still unable to reproduce the ultraviolet observations, which show
an excess of ultraviolet flux shortward of 1700~\AA~ and show profiles of the
strong Fe\,{\sc ii} and Mg\,{\sc ii} lines much weaker than the predicted ones.

Therefore, either the ATLAS12 model can still not predict the spectrum of
$\beta$ CrB, owing to the lack, in the code, of treatment of magnetic field 
effects and inhomogeneous chemical distributions, or we actually observe
the combined fluxes of $\beta$ CrB and of the unknown companion, 
which could well be a $\lambda$ Boo star, but also a much cooler star
affected by chromospheric emissions.

33~Lib shows a similar behaviour, but we have not been able as yet
to extend the sample to more stars, owing to a dramatic lack of IUE low 
resolution observations of Ap stars.

\begin{figure}[hbt]
\epsfysize=11cm
\epsfxsize=5.0cm
\hspace{0cm}\epsfbox{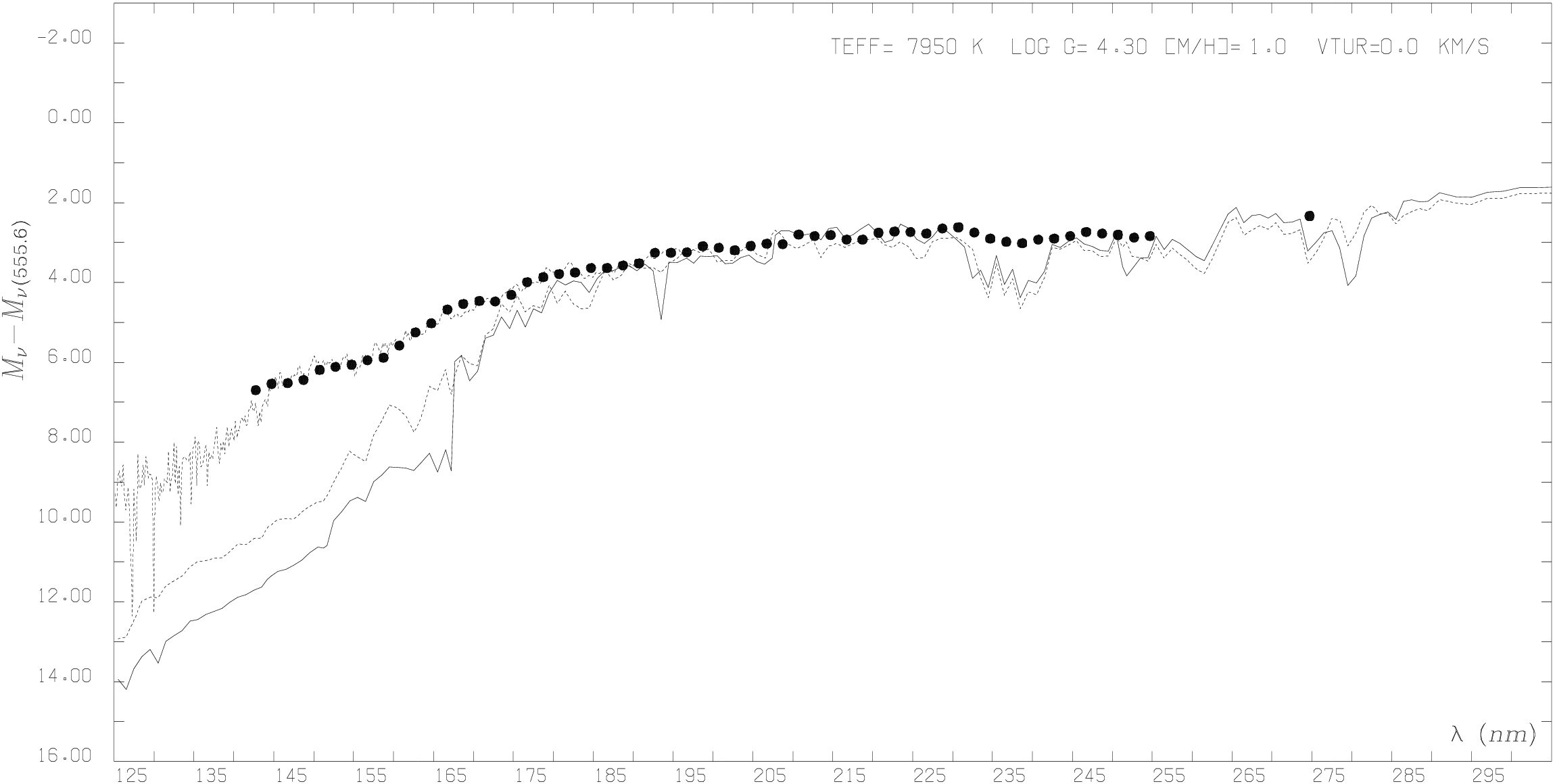}
\vspace{0.5cm}
\caption[h]{The ultraviolet energy distributions (in mag, relative to 5556~\AA)
from both ATLAS9 (full line) and ATLAS12 (dotted line) models are compared 
with the IUE  (dashed line) and TD1 S2/68 (dark points) observations}
\end{figure}

\end{document}